\documentstyle{article}

\def\abstract#1{\vskip 7mm 
        \begin{center}{\large Abstract}\par \smallskip
                \begin{minipage}[c]{12cm}
                        \small #1
                \end{minipage}
        \end{center}
}
\def\title#1{\begin{center}{\Large\bf #1}\end{center}}
\def\author#1{\vskip 5mm \begin{center}{#1}\end{center}}
\def\address#1{\begin{center}{\it #1}\end{center}}
\newread\epsffilein    
\newif\ifepsffileok    
\newif\ifepsfbbfound   
\newif\ifepsfverbose   
\newdimen\epsfxsize    
\newdimen\epsfysize    
\newdimen\epsftsize    
\newdimen\epsfrsize    
\newdimen\epsftmp      
\newdimen\pspoints     
\def\epsfbox#1{\global\def\epsfllx{72}\global\def\epsflly{72}%
   \global\def\epsfurx{540}\global\def\epsfury{720}%
   \def\lbracket{[}\def\testit{#1}\ifx\testit\lbracket
   \let\next=\epsfgetlitbb\else\let\next=\epsfnormal\fi\next{#1}}%
\def\epsfgetlitbb#1#2 #3 #4 #5]#6{\epsfgrab #2 #3 #4 #5 .\\%
   \epsfsetgraph{#6}}%
\def\epsfnormal#1{\epsfgetbb{#1}\epsfsetgraph{#1}}%
\def\epsfgetbb#1{%
\openin\epsffilein=#1
\ifeof\epsffilein\errmessage{I couldn't open #1, will ignore it}\else
   {\epsffileoktrue \chardef\other=12
    \def\do##1{\catcode`##1=\other}\dospecials \catcode`\ =10
    \loop
       \read\epsffilein to \epsffileline
       \ifeof\epsffilein\epsffileokfalse\else
          \expandafter\epsfaux\epsffileline:. \\%
       \fi
   \ifepsffileok\repeat
   \ifepsfbbfound\else
    \ifepsfverbose\message{No bounding box comment in #1; using defaults}\fi\fi
   }\closein\epsffilein\fi}%
\def\epsfsetgraph#1{%
   \epsfrsize=\epsfury\pspoints
   \advance\epsfrsize by-\epsflly\pspoints
   \epsftsize=\epsfurx\pspoints
   \advance\epsftsize by-\epsfllx\pspoints
   \epsfxsize\epsfsize\epsftsize\epsfrsize
   \ifnum\epsfxsize=0 \ifnum\epsfysize=0
      \epsfxsize=\epsftsize \epsfysize=\epsfrsize
     \else\epsftmp=\epsftsize \divide\epsftmp\epsfrsize
       \epsfxsize=\epsfysize \multiply\epsfxsize\epsftmp
       \multiply\epsftmp\epsfrsize \advance\epsftsize-\epsftmp
       \epsftmp=\epsfysize
       \loop \advance\epsftsize\epsftsize \divide\epsftmp 2
       \ifnum\epsftmp>0
          \ifnum\epsftsize<\epsfrsize\else
             \advance\epsftsize-\epsfrsize \advance\epsfxsize\epsftmp \fi
       \repeat
     \fi
   \else\epsftmp=\epsfrsize \divide\epsftmp\epsftsize
     \epsfysize=\epsfxsize \multiply\epsfysize\epsftmp   
     \multiply\epsftmp\epsftsize \advance\epsfrsize-\epsftmp
     \epsftmp=\epsfxsize
     \loop \advance\epsfrsize\epsfrsize \divide\epsftmp 2
     \ifnum\epsftmp>0
        \ifnum\epsfrsize<\epsftsize\else
           \advance\epsfrsize-\epsftsize \advance\epsfysize\epsftmp \fi
     \repeat     
   \fi
   \ifepsfverbose\message{#1: width=\the\epsfxsize, height=\the\epsfysize}\fi
   \epsftmp=10\epsfxsize \divide\epsftmp\pspoints
   \vbox to\epsfysize{\vfil\hbox to\epsfxsize{%
      \includegraphics{#1}%
      \hfil}}%
\epsfxsize=0pt\epsfysize=0pt}%

{\catcode`\%=12 \global\let\epsfpercent=
\long\def\epsfaux#1#2:#3\\{\ifx#1\epsfpercent
   \def\testit{#2}\ifx\testit\epsfbblit
      \epsfgrab #3 . . . \\%
      \epsffileokfalse
      \global\epsfbbfoundtrue
   \fi\else\ifx#1\par\else\epsffileokfalse\fi\fi}%
\def\epsfgrab #1 #2 #3 #4 #5\\{%
   \global\def\epsfllx{#1}\ifx\epsfllx\empty
      \epsfgrab #2 #3 #4 #5 .\\\else
   \global\def\epsflly{#2}%
   \global\def\epsfurx{#3}\global\def\epsfury{#4}\fi}%
\def\epsfsize#1#2{\epsfxsize}
\makeatletter
\@ifundefined{lesssim}{}{}
\@ifundefined{gtrsim}{}{}
\def\vereq#1#2{\lower3pt\vbox{\baselineskip1.5pt \lineskip1.5pt
\ialign{$\m@th#1\hfill##\hfil$\crcr#2\crcr\sim\crcr}}}
\makeatother

\begin{document}

\title{%
  Comparing Quantum Black Holes and Naked Singularities\footnote{Based on a talk given at JGRG10, Osaka,
September, 2000. To appear in the conference proceedings.}}
  
\author{%
  T. P. Singh}
  
\address{%
  Tata Institute of Fundamental Research, Homi Bhabha Road, Mumbai 400005, 
  India\\
  Yukawa Institute for Theoretical Physics, Kyoto University, 
  Kyoto 606--8502, Japan\\
  e-mail: tpsingh@nagaum.tifr.res.in}
\abstract{
  There are models of gravitational collapse in classical general relativity
which admit the formation of naked singularities as well as black holes.
These include fluid models as well as models with scalar fields as matter.
Even if fluid models were to be regarded as unphysical in their matter
content, the remaining class of models (based on scalar fields) generically
admit the formation of visible regions of finite but arbitrarily high curvature. Hence
it is of interest to ask, from the point of view of astrophysics, as to what
a stellar collapse leading to a naked singularity (or to a visible region of
very high curvature) will look like, to a far away observer. The emission of
energy during such a process may be divided into three phases - (i) the
classical phase, during which matter and gravity can both be treated
according to the laws of classical physics, (ii) the semiclassical phase,
when gravity is treated classically but matter behaves as a quantum field,
and (iii) the quantum gravitational phase. In this review, we first give a
summary of the status of naked singularities in classical relativity, and
then report some recent results comparing the semiclassical phase of black
holes with the semiclassical phase of spherical collapse leading to a naked
singularity. In particular, we ask how the quantum particle creation during
the collapse leading to a naked singularity compares with the Hawking
radiation from a star collapsing to form a black hole. It turns out that
there is a fundamental difference between the two cases. A spherical naked
star emits only about one Planck energy during its semiclassical phase, and
the further evolution can only be determined by the laws of quantum gravity.
This contrasts with the semiclassical evaporation of a black hole wherein
gravity can be treated classically all the way till the final stages of
evaporation until a Planck mass remnant remains. Hence spherical collapse
leading to a naked singularity provides an interesting and promising system
for testing our understanding of quantum gravity. The results on the
semiclassical phase of naked collapse reviewed here have been obtained in
collaboration with Barve, Harada, Iguchi, Nakao, Tanaka, Vaz and Witten
}

\section{Cosmic Censorship and Classical General Relativity}

Put in broad physical terms, the Cosmic Censorship Hypothesis states that
{\it generic} gravitational collapse of {\it physically reasonable} matter 
does not end in the formation of a {\it naked singularity}. A naked singularity
is a singularity which is visible to a far away observer, i.e. outgoing light rays
starting from the singularity terminate on the singularity in the past. By
`physically reasonable' we mean matter which satisfies one or more positive
energy conditions - matter which can in principle be prepared in the laboratory.
By `generic' we mean that the initial conditions leading to the formation of
the naked singularity are not special (of zero measure).

The Censorship Hypothesis is important in classical general relativity
because there are theorems, for instance the black hole area theorem, whose
proof assumes the validity of the Hypothesis. However, it is not obvious a priori
that the results of these theorems cannot be proved without assuming Censorship. 
This by itself is an interesting direction for research in the classical theory.

The issue of naked singularities is not trivialized by quantum gravity, even 
though it might be true that quantum gravity will remove the singularities of
classical general relativity, by replacing them by a smeared out region of 
Planck scale curvature. If the classical singularity resides inside an 
astrophysical black hole, such a smeared region will be invisible to a distant
observer. However, if the classical singularity is naked, the Planck curvature
region by which it will be replaced will be visible to the distant observer and 
will have physical consequences very different from that of the black hole. For
instance, it could give rise to an intense particle production which could differ
in character from the Hawking radiation from a black hole.

If we think of classical general relativity as a limit of quantum gravity, then we
should think of Cosmic Censorship in a broader perspective, and take the 
Hypothesis to mean: `Effects of quantum gravity cannot be observed in 
gravitational collapse'.
Looked at in this way, it would be more useful for physics if Censorship were to
be violated, as then the Universe could consist of naked stars - stars which 
according to the classical theory end as naked singularities, and in whose 
collapse we may witness the effects of quantum gravity.

Over the last twenty years or so, there have been many investigations of Cosmic
Censorship, largely using spherical models of gravitational collapse of physically
reasonable matter. The simplest of these models is dust collapse, which has been
studied in great detail and it has been shown that both black holes
and naked singularities form from generic initial conditions. Eardley and
Smarr \cite{eardley} used numerical methods to study the dust model.
Christodoulou \cite{chrisdust} analysed dust collapse with $C^{\infty}$ initial
data and Newman \cite{newman} showed that the consequent naked singularities 
were gravitationally weak. Dwivedi and Joshi developed a roots analysis to check naked
singularity formation \cite{djdust} and demonstrated the occurrence of strong
curvature naked singularities in dust collapse. Waugh and Lake \cite{lake}
showed the occurrence of strong curvature naked singularity in self-similar
dust collapse. The initial data leading to
these strong curvature naked singularities was worked out in \cite{jossin}
and a unified derivation of strong and weak naked singularities was obtained
in \cite{sinjos}. The structure of the apparent horizon in dust collapse was
studied in \cite{jjs} and a simplified derivation of the dust naked singularity
was given in \cite{bsvwdust}. Harada et al. \cite{hard1,hard2} investigated
the stability of spherical dust collapse against non-spherical perturbations.
The strength of the singularity has been studied by \cite{desh}, the nature of
non-spacelike geodesics in \cite{desh2} and the role of initial data in 
\cite{id,tavakol}. A cosmological constant has been included in \cite{lamb1,lamb2}.

Similar results have been obtained for null dust collapse described by
the Vaidya model \cite{null1,null2,null3,null4,null5,null6,null7,null8,
null9,null10,null11,null12}.

The next class of models is spherical perfect fluid collapse, which has been
examined numerically \cite{oripiran} and analytically \cite{djp1, djp2} with and 
without the assumption of
self-similarity \cite{harp, coop} and again it has been found that both black hole and naked
singular solutions arise from generic initial data. 

Another tractable class of models is the spherical collapse of fluids which have
only tangential pressure, and for which the radial pressure is assumed to be
negligible. For these models also it is observed that covered as well as naked
solutions exist \cite{tang1,tang2,tang3,tang4,tang5,tang6,tang7}.

One might consider dividing physically reasonable matter models into two 
classes - fluids and fields. The former may be regarded as a phenomenological
description of matter while the latter may be regarded as fundamental. Could it
be that naked singularities do not arise generically in the collapse of matter which is
treated as a field? The work of Christodoulou 
\cite{chris1,chris2,chris3,chris4,chris5,chris6,chris7} on massless scalar 
fields suggests this could be the case. It should be said though that this issue
of phenomenological versus fundamental needs further investigation, and it is not
entirely clear why fluids should admit naked singularities but fields should not.

Even if one were to accept that Cosmic Censorship should be investigated with only
fundamental matter, there is a broader outlook which in my view seals the case
against Censorship. This has to do with the result that in spherical scalar
collapse one can form black holes of arbitrarily small mass \cite{chop}. Hence, 
although the naked singularity itself forms from non-generic initial data, 
visible regions of arbitrarily high curvature (the curvature on the horizon of 
the arbitrarily small black hole) form generically.

So it may turn out to be true that the collapse of fundamental matter does not
generically form naked singularities. Yet it appears to be more physical to think
of Cosmic Censorship as a statement not about singularities, but about regions of
extremely high curvature (approaching Planck scale) and in this sense censorship
does not appear to hold in spherical general relativity, even if only 
fundamental matter is considered. 

Attempts have also been made to study the nature of singularities in spherical
collapse without having to restrict to a specific form of matter
\cite{lakeg,djc,s1,s2,nolan}

If we accept that Censorship does not hold, in the sense described in the previous
paragraph, are there any interesting implications for astrophysics? This is far
from clear. We have to address the following questions: if a star collapses to 
form a naked singularity (or a black hole of very small mass) what does the 
process look like to a far off observer? How much energy is emitted and over what
duration and with what spectrum? How does the back reaction from the emission
modify the collapse? In this paper, we address some of these questions.

Let us broadly divide `censorship violating' spherical collapse into two 
classes: one which results in a naked singularity at the center, and the second 
which results in a black hole of very small mass while the rest of the mass 
disperses to infinity. In this article we focus attention on the first of these 
two classes.

The emission from a star collapsing to form a naked singularity may be divided into
three phases - classical, semiclassical and quantum gravitational. The classical
phase would consist of the radiation which is emitted while both matter and gravity
can be treated by the laws of classical physics. Could most of the star blow itself 
up during this phase? We do not know for sure. This phase has been investigated by 
Joshi, Dadhich and Maartens \cite{maartens}. In the next phase of the collapse, matter will have
to be treated as a quantum field whereas gravity can be treated classically (the
semiclassical phase). This is the analog of the black hole phase in which the black
hole emits Hawking radiation. The final collapse phase is quantum gravitational.
 
The main purpose of the present article is to report some recent results 
of Harada et al. \cite{harada} on the 
semiclassical phase of a spherical star forming a naked singularity. We show
that the nature of particle creation in this phase is very different from the
Hawking radiation from a black hole and that quantum gravity is essential for
understanding the gravitational collapse to a naked singularity.

Some of the earlier reviews on Cosmic Censorship are listed in 
\cite{rev1,rev2,rev3,rev4,rev5,rev6,rev7,rev8}. 
The reader might find it useful to read the present article in conjunction with
some of the earlier reviews.

\section{Quantum Black Holes Versus Quantum Naked Singularities}

Consider a spherical star which collapses to form a black hole and another
spherical star which collapses to form a naked singularity. Let us quantize
a massless scalar field on the background of either star, and prepare the
field to be initially in the Minkowski vacuum, when the gravitational field
of the star is weak enough so that spacetime is nearly flat. As is well known,
the star which settles into the black hole state emits Hawking radiation, which
has a thermal spectrum with temperature inversely proportional to the mass of the
black hole.

We are interested in finding out the nature of the corresponding quantum radiation
emitted in the naked singularity case. A comparison of the representative Penrose
diagrams (Fig. 1) in the two cases

\begin{center}
\leavevmode\epsfxsize=5in\epsfbox{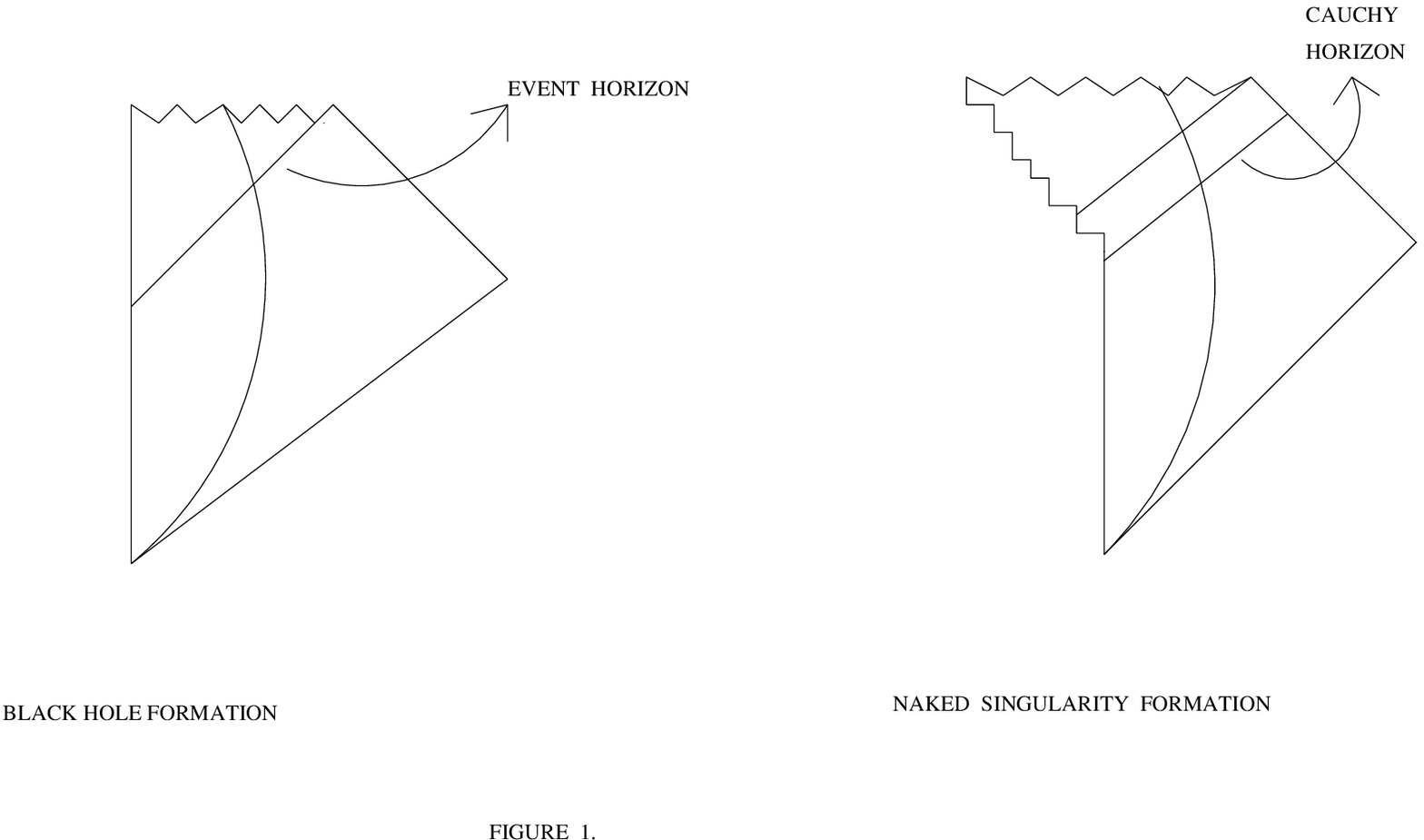}
\end{center}

\noindent shows the immediate obstacle to repeating 
Hawking's Bogoliubov transformation calculation in the naked case. In the latter
case, a part of the asymptotic future null infinity is exposed to the naked null
singularity, making it impossible to define a complete set of modes on 
${\cal I}^{+}$ unless some boundary conditions are prescribed on the naked
singularity. Hence it is not possible to carry out the Bogoliubov transformations
and a particle creation calculation the way it is done for the black hole case.

However, there is in principle no problem in calculating the expectation value of
the quantum stress energy tensor of the scalar field in the region of the 
Penrose diagram prior to the Cauchy horizon.  From this, one can obtain the
expectation value of the outgoing quantum flux everywhere in that part of 
spacetime which is not exposed to the naked singularity, and in particular on that
part of ${\cal I}^{+}$ which is prior to its intersection with the Cauchy horizon.
Understandably, for the black hole case, such a calculation yields the standard
Hawking thermal flux. We are now asking for the result of this calculation in the
case when the star collapses to a naked singularity.

In the case of a four-dimensional spherical collapse Ford and Parker \cite{ford}
obtained a formula for the outgoing quantum flux on the Cauchy horizon
in the geometric optics
approximation, using point-splitting regularisation. Suppose an ingoing ray
$V=constant$ coming in from ${\cal I}^{-}$ gets mapped into an outgoing ray
$U={\cal F}(V)$ when it reaches $\cal{I}^{+}$. Then the outgoing quantum flux,
for a given low angular momentum mode, is
\begin{equation}
P={\hbar\over 24\pi} \left [ {{\cal F}'''\over {\cal F}'^{3}} -
{3\over 2}\left({{\cal F}''\over {\cal F}'^{2}}\right)^{2}\right].
\end{equation}  
In terms of the inverse map $V={\cal G}(U)$ the power is given by
\begin{equation}
P={\hbar\over 24\pi} \left [{3\over 2}\left({{\cal G}''\over {\cal G}'^{2}}\right)^{2}
 -{{\cal G}'''\over {\cal G}'} 
\right].
\label{power}
\end{equation}

For a two-dimensional spacetime model, the outgoing quantum flux can be
calculated exactly. Suppose the spacetime is given by double null
coordinates $u,v$ as
\begin{equation}
ds^{2}=C^{2}(u,v) du dv.
\end{equation}
Then, the use of the trace anomaly and the divergencelessness condition
\begin{equation}
Tr<in|T_{\mu\nu}|in>={{\cal R}\over 24\pi},\hskip 0.2 in <T^{\mu\nu}>_{;\nu}=0
\end{equation}
allows the outgoing flux to be written as \cite{fulling}
\begin{equation}
<T_{uu}>=-{1\over 12\pi} C {\partial^{2}\over \partial u^{2}}\left(1/C\right).
\label{twod}
\end{equation}
Here, ${\cal R}$ is the curvature scalar for the two dimensional spacetime.
In the four dimensional case as well as for the 2-d approximation, one needs
to be able to solve for null rays in order to calculate the flux. We now discuss
these two approximations one by one. Before doing that, we give a brief outline
of the naked singularity in the classical spherical dust collapse model
\cite{bsvwdust}.

\vskip 0.2 in

\centerline{\bf Spherical Dust Collapse}
\smallskip
In comoving coordinates $(t,r,\theta ,\phi )$ the spacetime metric for
marginally bound spherical dust collapse is given by 
\begin{equation}
\label{met}ds^2=dt^2-R^{\prime 2}dr^2-R^2d\Omega ^2 
\end{equation}
where $R(t,r)$ is the area radius at time $t$ of the shell having the
comoving coordinate $r$. A prime denotes partial derivative w.r.t. $r$. The
energy-momentum tensor for dust has only one non-zero component $%
T_0^0=\epsilon (t,r)$, which is the energy density. The Einstein equations
for the collapsing cloud are 
\begin{equation}
\label{e1}\frac{8\pi G}{c^4}\epsilon (t,r)=\frac{F^{\prime }}{R^2R^{\prime }}%
,\;\;\dot R^2=\frac{F(r)}R. 
\end{equation}
A dot denotes partial derivative w.r.t. time $t$. The function $F(r)$
results from the integration of the second order equations. Henceforth we
shall set $8\pi G/c^4=1$.

The second of these equations can be easily solved to get 
\begin{equation}
\label{R}R^{3/2}(t,r)=r^{3/2}-\frac 32\sqrt{F}t 
\end{equation}
where we have used the freedom in the scaling of the comoving coordinate $r$
to set $R(0,r)=r$ at the starting epoch of collapse, $t=0$. It follows from
the first equation in (\ref{e1}) that the function $F(r)$ gets fixed once
the initial density distribution $\epsilon (0,r)=\rho (r)$ is given, i.e. 
\begin{equation}
\label{eff}F(r)=\int \rho (r)r^2dr. 
\end{equation}
Hence $F(r)$ has the interpretation of being twice the mass to the interior
of the shell labeled $r$. If the initial density $\rho (r)$ has a series
expansion 
\begin{equation}
\label{de}\rho (r)=\rho _0+\rho _1r+\frac 1{2!}\rho _2r^2+\frac 1{3!}\rho
_3r^3+... 
\end{equation}
near the center $r=0$, the resulting series expansion for the mass function $%
F(r)$ is 
\begin{equation}
\label{ma}F(r)=F_0r^3+F_1r^4+F_2r^5+F_3r^6+... 
\end{equation}
where $F_q=\rho _q/q!(q+3)$, and $q=0,1,2,3..$We note that we could set $%
\rho _1=0$ without in any way affecting the conclusions of this paper.
Further, the first non-vanishing derivative in the series expansion in (\ref
{de}) should be negative, as we will consider only density functions which
decrease as one moves out from the center.

According to (\ref{R}) the area radius of the shell $r$ shrinks to zero at
the time $t_c(r)$ given by 
\begin{equation}
\label{tc}t_c(r)=\frac{2r^{3/2}}{3\sqrt{F(r)}}. 
\end{equation}
At $t=t_c(r)$ the Kretschmann scalar 
\begin{equation}
\label{kr}K=12\frac{F^{\prime 2}}{R^4R^{\prime 2}}-32\frac{FF^{\prime }}{%
R^5R^{\prime }}+48\frac{F^2}{R^6} 
\end{equation}
diverges at the shell labeled $r$ and hence this represents the formation of
a curvature singularity at $r.$ In particular, the central singularity, i.e.
the one at $r=0$, forms at the time 
\begin{equation}
\label{sing}t_0=\frac 2{3\sqrt{F_0}}=\frac 2{\sqrt{3\rho _0}}. 
\end{equation}
At $t=t_0$ the Kretschmann scalar diverges at $r=0$. Near $r=0,$ we can
expand $F(r)$ and approximately write for the singularity curve 
\begin{equation}
\label{texp}t_c(r)=t_0-\frac{F_n}{3F_0^{3/2}}r^n. 
\end{equation}
Here, $F_n$ is the first non-vanishing term beyond $F_0$ in the expansion (%
\ref{ma}). We note that $t_c(r)>t_0$, since $F_n$ is negative.

We wish to investigate if the singularity at $t=t_0,r=0$ is naked, i.e. are
there one or more outgoing null geodesics which terminate in the past at the
central singularity. We restrict attention to radial null geodesics. Let us
start by assuming that one or more such geodesics exist, and then checking
if this assumption is correct. Let us take the geodesic to have the form 
\begin{equation}
\label{geo}t=t_0+ar^\alpha 
\end{equation}
to leading order, in the $t-r$ plane, where $a>0$, $\alpha >0$. In order for
this geodesic to lie in the spacetime, we conclude by comparing with (\ref
{texp}) that $\alpha \geq n$, and in addition, if $\alpha =n$, then $%
a<-F_n/3F_0^{3/2}$.

As is evident from the form (\ref{met}) of the metric, an outgoing null
geodesic must satisfy the equation 
\begin{equation}
\label{ng}\frac{dt}{dr}=R^{\prime }. 
\end{equation}
In order to calculate $R^{\prime }$ near $r=0$ we first write the solution (%
\ref{R}) with only the leading term $F_n$ retained in $F(r)$ in (\ref{ma}).
This gives 
\begin{equation}
\label{Ra}R=r\left( 1-\frac 32\sqrt{F_0}\left[ 1+\frac{F_n}{2F_0}r^n\right]
\,t\right) ^{2/3}. 
\end{equation}
Differentiating this w.r.t. $r$ gives 
\begin{equation}
\label{Rp}R^{\prime }=\left( 1-\frac 32\sqrt{F_0}\left[ 1+\frac{F_n}{2F_0}%
r^n\right] \,t\right) ^{-1/3}\,\,\left( 1-\frac 32\sqrt{F_0}t-\frac{(2n+3)F_n%
}{4\sqrt{F_0}}r^nt\right) . 
\end{equation}

Along the assumed geodesic, $t$ is given by (\ref{geo}). Substituting this
in $R^{\prime }$ and equating the resulting $R^{\prime }$ to $dt/dr=\alpha
ar^{\alpha -1}$ gives 
\begin{equation}
\label{maj}\alpha ar^{\alpha -1}=\frac{\,\left( 1-\frac 32\sqrt{F_0}\left[
t_0+ar^\alpha \right] -\frac{(2n+3)F_n}{4\sqrt{F_0}}r^n\left[ t_0+ar^\alpha
\right] \right) }{\left( 1-\frac 32\sqrt{F_0}\left[ 1+\frac{F_n}{2F_0}%
r^n\right] \,\left[ t_0+ar^\alpha \right] \right) ^{1/3}}\,. 
\end{equation}
$\,$ This is the key equation. If it admits a self-consistent solution then
the singularity will be naked (i.e. at least one outgoing null geodesic will
terminate at the singularity), otherwise not. We simplify this equation by
putting in the requirement mentioned earlier, that $\alpha \geq n$. Consider
first $\alpha >n$. In this case we get, to leading order 
\begin{equation}
\label{oo}\alpha ar^{\alpha -1}=\left( 1+\frac{2n}3\right) \left( -\frac{F_n%
}{2F_0}\right) ^{2/3}r^{2n/3} 
\end{equation}
which implies that $\alpha =1+2n/3$, and $a=(-F_n/2F_0)^{2/3}$. By
substituting integral values for $n$ we find that only for $n=1$ and $n=2$
the condition $\alpha >n$ is satisfied. Hence the singularity is naked for $%
n=1$ and $n=2$, i.e. for the models $\rho _1<0$ and for $\rho _1=0,\rho _2<0 
$. There is at least one outgoing geodesic given by (\ref{geo}) which
terminates in the central singularity in the past. If $n>3$ then the
condition $\alpha >n$ cannot be satisfied and the singularity is not naked.
This is the case $\rho _1=\rho _2=\rho _3=0$. It includes as a special
case the homogeneous black hole model of Oppenheimer and Snyder.

Consider next that $\alpha =n$. In this case we get from (\ref{maj}) that 
\begin{equation}
\label{n3}nar^{n-1}=\frac{-\frac 32a\sqrt{F_0}-\frac{(2n+3)F_n}{6\sqrt{F_0}}%
}{\left( -\frac{F_n}{2F_0}-\frac{3a}2\sqrt{F_0}\right) ^{1/3}}r^{2n/3} 
\end{equation}
which implies that $n=3$ and gives an implicit expression for $a$ in terms
of $F_3$ and $F_0$. This expression for $a$ can be simplified to get the
following quartic for $a$: 
\begin{equation}
\label{qu3}12\sqrt{F_0}%
a^4-a^3(-4F_3/F_0+F_0^{3/2})-3F_3a^2-3F_3^2/F_0^{3/2}a-(F_3/F_0)^3=0. 
\end{equation}
By defining $b=a/F_0$ and $\xi =F_3/F_0^{5/2}$ this quartic can be written
as 
\begin{equation}
\label{qu4}4b^3(3b+\xi )-(b+\xi )^3=0. 
\end{equation}
The singularity will be naked if this equation admits one or more positive
roots for $b$ which satisfy the constraint $b<-\xi /3$. This last inequality
is the same as the condition $a<-F_n/3F_0^{3/2}$ given below equation (\ref
{geo}). We note that $\xi $ is negative. This quartic can be made amenable
to further analysis by substituting $Y=-2b/\xi ,$ and then $\eta
=-1/6\xi $, so as to get 
\begin{equation}
\label{qn2}Y^3(Y-2/3)-\eta (Y-2)^3=0. 
\end{equation}
This quartic has two positive real roots provided $%
\eta \geq \eta _1$ or $\eta \leq \eta _2$ where 
\begin{equation}
\label{al}\eta _1=\frac{26}3+5\sqrt{3},\;\eta _2=\frac{26}3-5\sqrt{3}.\; 
\end{equation}
We also require that $Y<2/3$. By examining the quartic (\ref{qn2}) one can
see that if $\eta \geq \eta _1$ then $Y\geq 2$; hence this range of $\eta $
is ruled out. Thus the singularity is naked provided $\eta \leq \eta _2$, or
equivalently $\xi \leq -25.9904$.

The above results give the conditions on the initial data for which the
singularity is at least locally naked, i.e. light manages to escape
from the singularity. Further constraints have to be put on the initial
distribution in order to obtain a globally naked singularity, i.e. one in
which light rays starting from the singularity escape the dust cloud entirely.
In general a choice of initial conditions is available for which a singularity
which is at least locally naked is also globally naked. The spacetime with
a globally naked singularity is described by the Penrose diagram in 
Figure 1. From the point of view of the present article the globally
naked singularities are the interesting ones; they are the ones whose
asymptotic quantum appearance differs from those of black holes.

A useful special case of the class $\alpha=n=3$ is the self-similar model,
one which admits a homothetic Killing vector field \cite{oripiran}. The
condition for nakedness is the same as for the whole class $\alpha=n=3$ and
the singularity is necessarily globally naked.

We recall next the quantisation of a massless scalar field on the background
spacetime provided by dust collapse.

\vskip 0.2 in

\centerline{\bf Geometric Optics Approximation}
\smallskip
We now consider the application of the Ford-Parker formula (\ref{power})
for calculating the quantum flux of a massless scalar field on the
dust background.
For the case in which the collapse ends in a black hole, the map 
${\cal G}(U)$ is given by
\begin{equation} 
{\cal G}(U)= A - B \exp(-U/4M).
\end{equation}
Using this map in (\ref{power}) gives the expected thermal Hawking flux
\begin{equation}
P={1\over 768\pi M^{2}}.
\label{haw}
\end{equation}
The Hawking flux for the black hole is found using the map ${\cal F}(v)$ which in turn
depends on the behaviour of null rays in the vicinity of the event horizon. In this
sense the map is universal for black holes - it does not depend on the interior
geometry of the star.

On the other hand the map ${\cal G}(U)$ very much depends on the internal geometry in 
the naked case. Hence one has to first specify the internal solution and calculate
the propagation of null rays inside the star. There are only a limited number of cases
where this has been done so far. The first example to be considered was that of
shell-crossing dust naked singularity \cite{ford} for which the outgoing flux was calculated
and shown to be finite.

Of greater interest is the shell-focusing naked central singularity in spherical dust
collapse, described above. This was considered by Barve et al.
\cite{barve1} for the self-similar dust collapse and by Singh and Vaz \cite{singh1}
for self-similar null dust collapse. The map ${\cal G}(U)$ was calculated and shown
to be
\begin{equation}
{\cal G}(U) = A - B(V_{0}-V)^{\gamma}
\end{equation}
where $\gamma$ is a positive constant determined by the collapse model.
The flux can be calculated using (\ref{power}) and near the Cauchy horizon
$U=U_{0}$ it behaves as
\begin{equation}
P(U)\sim {\hbar\over (U_{0}-U)^{2}}.
\label{div}
\end{equation}
It thus diverges in the approach to the Cauchy horizon. Such a divergence appears to be a
universal feature of collapse leading to a naked singularity. It has also been observed 
in dust collapse models which do not assume self-similarity \cite{harada2,harada3}, for which the 
divergence is of the form
\begin{equation}
P(U)\sim {\hbar\over (U_{0}-U)^{3/2}}.
\end{equation}
What is the meaning of this divergence? At first sight, it suggests that as a result of 
the intense particle creation accompanying the formation of the naked singularity, a
huge amount of energy is emitted - essentially the whole star blows up. However, we
realize that in inferring this divergence we have extended the semiclassical approximation
all the way up to the epoch of naked singularity formation, when the curvature of the
background spacetime diverges. But we should not a priori trust the results of the
semiclassical theory beyond the epoch when the curvature in the central region of
the star approaches Planck scale. This happens about one Planck time before the singular
epoch, and this can be shown \cite{harada} to translate to $U_{0}-U\approx t_{Planck}$ in
(\ref{div}). It then follows from (\ref{div}) that during the domain of validity of the
semiclassical approximation the emitted energy is only about one Planck energy, even if the
total initial mass of the star is much larger, say a solar mass. Hence this system is  very
different from the black hole, for which essentially the entire mass of the collapsing object
evaporates during the semiclassical phase and Planck scale curvatures are realised only when
a Planck mass remnant remains.

The result that the semiclassical energy emission is very small appears to be supported
by the observation that the map ${\cal F}(V)$ in the naked case is determined by
examining a small spacetime region near the central singularity \cite{tanaka}. 

We mentioned earlier that the calculation of the spectrum of created particles,
which is based on Bogoliubov transformations, cannot be performed in the naked
case without specifying boundary conditions on the singularity. If the semiclassical
approximation were assumed to be valid all the way up to the formation of the
naked singularity, the resulting divergence on the Cauchy horizon would suggest
a complete evaporation of the star. Under such a situation it appears reasonable to
assume that the singularity does not form in the first place, and an appropriate
boundary condition would be that of no emission from the null singularity. Under
these assumptions it has been shown that the emitted spectrum is non-thermal
\cite{spec1,spec2,harada3}. It however remains to be understood how these results on the
spectrum are modified by the quantum gravity phase which comes into play about
a Planck time prior to the singularity.

\vskip 0.2 in

\centerline{\bf 2-d collapse models}

\smallskip

The 2-d models that we consider are obtained by suppressing the angular coordinates
in a 4-d spherical model.
If one applies the 2-d quantum flux result to the Schwarzschild geometry, one again 
obtains the Hawking result (\ref{haw}) for the power. 
The first application to a naked singularity model was by Hiscock et al. 
\cite{null1} who
calculated the outgoing flux for the naked singularity in self-similar
Vaidya null dust collapse and showed it to diverge on the Cauchy horizon.
A similar divergence was found by Barve et al. \cite{barve2} for the case
of self-similar Tolman-Bondi dust collapse:
\begin{equation}
<T_{uu}>\sim {\hbar\over (z-z_{-})^{2}}
\end{equation}
where $z$ is the self-similarity parameter and $z=z_{-}$ is the Cauchy
horizon. It was shown in \cite{barve3} that the divergence on the
Cauchy horizon persists in non-self-similar 2-d dust collapse.

Once again, it is apparent that the semiclassical evolution should be
stopped about a Planck time before the singular epoch. It then turns
out that in the 2-d model as well the energy emitted during the
semiclassical phase is about one Planck energy.

An important aspect of the above analysis is that we have ignored the
back-reaction, and one could well ask if our conclusions could be
affected by its inclusion. We now give arguments as to why 
the back reaction cannot be important during the semiclassical evolution,
in the present case. One could propose two different criteria for
deciding as to when the back reaction becomes important. One is that
the total flux received at infinity becomes comparable to the mass of
the collapsing star. As we have seen above, if the mass of the
collapsing star is much greater than the Planck mass (as is of course
usually the case) then the back reaction does not become important during
the semiclassical phase.

The second criterion could be that the energy density of the quantum
field becomes comparable to the background energy density, {\it
inside} the star. In a four-dimensional model, it is difficult to
establish whether this happens. However, one can study the evolution
of the quantum field inside the star using the 2-d model obtained by
suppressing the angular part of the 4-d Tolman-Bondi model 
\cite{barve2}.  Using this
model, Iguchi and Harada \cite{iguchi} have shown that the back reaction
does not become
significant during the semiclassical evolution, inside the star. It is
plausible that this result holds for the four dimensional stellar dust
model as well, in which case we can conclude that quantum
gravitational effects are more important than the back reaction in
deciding the evolution of the star.

Again we see that the semiclassical evolution is very different from
the black hole case, since the back reaction plays a crucial role 
during the semiclassical history of the black hole.

\section{Discussion and Unresolved Issues}

We see that when a spherical dust star collapses to form a naked
singularity, the character of the semiclassical emission (resulting from
the quantization of a massless scalar field on the dust background) is
very different from the Hawking emission from a dust star collapsing to
form a black hole. The duration of the semiclassical phase for the naked
case is of the order of the collapse time $\sim 1/\sqrt{G\rho}$, in the
comoving frame as well as for the asymptotic observer. (Since there is
no event horizon preceding the formation of the singularity, there is
no significant time-dilation caused by the gravitational redshift on the
boundary). This time scale is of course extremely short, compared to the 
astronomical time scale for the evaporation of an astrophysical black hole. 

With hindsight, it is not difficult to see why the semiclassical phase
of a naked collapse is so different from that of collapse leading to
a black hole. The central naked singularity forms extremely quickly, as
seen by a far away observer, and this naked central spacetime region can
only be dealt with using quantum gravity. In contrast, the asymptotic
observer never gets to see the singularity inside the black hole, except
during the final stages of black hole evaporation. Hence the black hole
semiclassical epoch is extremely long.

Starting at about one Planck time before singularity formation, we must
analyse the further evolution of the naked star using the laws of the
(yet unknown) quantum theory of gravity. We thus have a novel situation where
a spherical dust star (say of one solar mass) enters the quantum gravity
phase with essentially all its mass intact, and unlike the black hole,
it does so very much within the lifetime of the Universe. This is perhaps
the first time that physicists have explicitly encountered a dynamical
system whose further evolution, within the age of the Universe, cannot
be understood without a knowledge of quantum gravity! For an alternative
discussion of Cosmic Censorship and quantum gravity see \cite{hod}.
Naked singularities in a dilaton gravity context are discussed in \cite{v1,v2}.

What happens to this spherical dust star next? How does the apparently
`quantum gravitational' region near the center interact with the
apparently classical regions which are away from the center? Does the
entire star evaporate explosively, as suggested by the semiclassical
theory, or does it settle into a black hole and then emit slowly by the
standard Hawking radiation? Is there a way in the quantum theory to 
separate the evolution of the central region from that of the outer
region? These are fascinating and challenging conceptual questions to 
which there are no obvious answers, and which perhaps cannot be addressed
without quantum gravity. Nonetheless, they provide a valuable testbed
for examining candidate theories of quantum gravity.

Motivated by such issues, Vaz et al. \cite{vaz} have begun an investigation
of quantum gravitational effects in spherical dust collapse leading to
a naked singularity. As a midisuperspace quantum gravity model, one sets
up a Wheeler-deWitt equation for the collapse, in terms of physically
well-motivated canonical variables. This approach is inspired by Kuchar's
canonical formalism for the Schwarzschild black hole \cite{kuchar}. By
investigating the solutions of the Wheeler-deWitt equation for this system
one hopes to address the questions raised in the previous paragraph.

As an aside, one could marvel at the richness of the dust collapse system,
whose studies were first initiated by Oppenheimer and Snyder, way back in
1939. Their work on homogeneous collapse was over the years generalised
to include shell-crossings and inhomogeneities, and then used as a background
model in quantum field theory in curved space, and now as a quantum gravity
model! We can be sure that we have not yet learnt all that dust collapse has
to teach us.

Does this picture of semiclassical naked collapse persist for spherical
naked singularities that form with other kinds of matter, say fluids?
We think it does, because the semiclassical properties are related to the
occurrence of the naked singularity in the small central region - this is
a property generic to the formation of a naked singularity in spherical
collapse. It is not restricted to the choice of dust as matter. Hence we
believe that the choice of dust as such is irrelevant to the importance
of quantum gravity in spherical naked collapse.

The other class of censorship violating spherical models which we mentioned
at the beginning of this article are those which form small mass black holes,
and for which the remaining mass escapes to infinity. What kind of results does
one expect for the semiclassical quantization of such systems? We would like
to conjecture that the semiclassical phase will be very similar to the dust
naked singularity system. The small mass black hole represents a region of
extremely high (Planck scale) curvature which cannot be dealt with semiclassically,
and must be analysed using quantum gravity. Furthermore, this region (unlike for
an evaporating black hole) is realised on a very short gravitational collapse
time scale. The only physical difference from the dust case is the dispersion
of the outer region, which does not constitute a significant difference in so
far as the semiclassical quantisation is concerned.

Hence we can say with confidence that the picture of semiclassical 
spherical collapse which we have seen in the dust model is representative
of censorship violating spherical collapse. The semiclassical phase of such
collapse emits an insignificant amount of energy and the full evolution can
only be understood in a quantum gravity framework.

Throughout the article, we have carefully emphasized that these semiclassical 
results apply only to naked singularities forming in spherical collapse. The
picture could well be very different if naked singularities were to form in
non-spherical collapse, for instance in the collapse to a elongated cigar shaped
naked singularity. Under such a situation an entire region (and not just a limited
central region) develops visible high curvature, and the particle creation and energy
release during the process could be enormous and explosive. This is an important 
outstanding problem - is the semiclassical phase of highly non-spherical
naked collapse similar to or very different from the semiclassical phase of naked
spherical collapse? Also, the nature of gravity wave emission during the formation of
aspherical naked singularities \cite{nakamura, thorne} is a promising 
research problem. Of course, it
should be added that our knowledge about formation of non-spherical naked 
singularities is at present much more restricted compared to spherical ones
\cite{nonsph1,nonsph2,nonsph3,nonsph4,nonsph5}.

Finally, one could well ask whether its worth anyone's effort to investigate
the possible occurrence and astrophysical properties of naked singularities.
An instructive answer can be found in the history of black hole physics.
Many decades ago, when a handful of relativists were trying to understand
the physical properties of black holes, very few astronomers and astrophysicists
took them seriously - and their efforts were often regarded as exercises in
mathematical physics, irrelevant to the real world of astronomical observations.
Needless to say, all this has changed with the detection of black holes in the
centers of galaxies and in x-ray binaries. But its fair to add that an understanding
of these detections would still be long in coming had relativists not prepared
the ground by working out what black holes would look like, if they were to exist.

With regard to research on naked singularities and naked stars, we are where black
hole physics was, say seventy years ago. Only further research can tell whether 
naked stars occur in nature. Considering the potential role of
such systems as experimental testbeds for quantum gravity it seems worth the
effort to explore their consequences. The possibility that gamma-ray bursts and/or some 
yet undiscovered astronomical object can be identified with naked stars cannot be ruled 
out at present \cite{grbs1,grbs2}. Even if the results of such searches turn out
to be negative, one will at least have found a route to proving Cosmic Censorship.     

\bigskip

\noindent {\bf Acknowledgments:} It is a pleasure to thank the Yukawa Institute
for Theoretical Physics, Kyoto for its warm hospitality and the organisers of
JGRG10 for a lively and fruitful meeting. I would also like to thank my
collaborators Sukratu Barve, Tomohiro Harada, Hideo Iguchi, Sanjay Jhingan, Pankaj Joshi,
Ken-ichi Nakao, Takahiro Tanaka, Cenalo Vaz and Louis Witten. I am grateful to
Hideo Kodama and Takashi Nakamura for their support and encouragement. 
I acknowledge the partial support of the 
Funda\c{c}\~ao para a Ci\^encia e a Tecnologia (FCT),
Portugal under contract number SAPIENS/32694/99.

\bigskip

\noindent {\bf A note on the references:} No claim is made that the
references are exhaustive. If there are papers references to which
have been inadvertently overlooked, I would like to apologise for
the same.

\end{document}